\documentclass[
    ,final            % use final for the camera ready runs
  ]
  {aipproc}
\usepackage{graphicx}
\usepackage{epsf}
\usepackage{psfig}
\usepackage{graphics}
\usepackage{lscape}
\usepackage{natbib}

\layoutstyle{6x9}

\begin{document}

\title{The origin of the relationship between black hole mass and host galaxy bulge luminosity}

%<Replace this text with PACS numbers; choose from this list:
%                \texttt{http://www.aip..org/pacs/index.html}>}

%Keywords should describe the main topics of the research reported in the article, with 3-8 keywords typically being sufficient.

\author{C. Martin Gaskell}{address={Astronomy Department, University of Texas, Austin, TX 78712-0259}}

\begin{abstract}
There is a strong decrease in scatter in the $M_\bullet$ -- $L_{bulge}$ relationship with increasing luminosity and very little scatter for the most luminous galaxies.  It is shown that this is a natural consequence of the substantial initial dispersion in the ratio of black hole mass to total stellar mass and of subsequent galaxy growth through hierarchical mergers.  ``Fine-tuning'' through feedback between black hole growth and bulge growth is neither necessary nor desirable.

\end{abstract}

\maketitle

The dispersions in the relationships between black hole mass, $M_\bullet$, host galaxy bulge luminosity, $L_{bulge}$, and stellar velocity dispersion have been shown \cite{Gaskell09,Gaskell10} to decrease strongly with increasing $L_{bulge}$ (see Fig. 1).  The trend in Fig. 1 can easily be modeled by assuming that bulges grow through mergers and that the $M_\bullet / L_{stars}$ ratio initially has a log-normal distribution with a substantial dispersion as is observed for the lowest luminosity galaxies.  A substantial scatter in the initial $M_\bullet / L_{stars}$ ratio is required. ``Fine-tuning'' through feedback is unnecessary and produces too low a dispersion in $M_\bullet / L_{bulge}$.

\vspace{0.3cm}
\begin{center}
\includegraphics[height=.25\textheight]{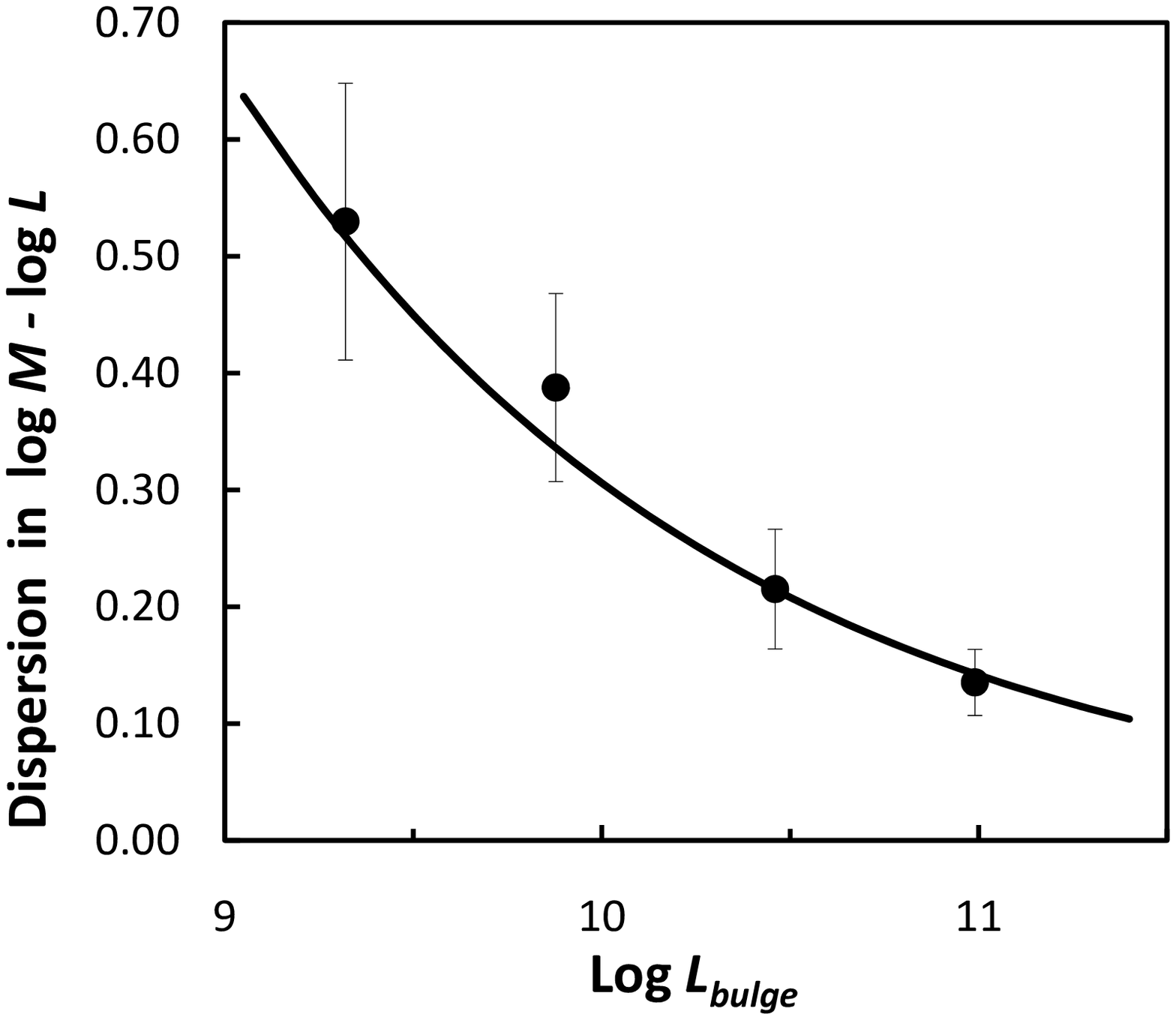}
\end{center}

\noindent {\footnotesize {\bf FIGURE 1.} ~~The 1-$\sigma$ scatter in the AGN $M_\bullet$ -- $L_{bulge}$ relationship as a function of bulge luminosity.  Observed points from \cite{Gaskell10}.  The solid line shows the predicted decrease in scatter for hierarchical merging.

\end{document}